\documentstyle[11pt]{article} 
\newcommand{\re}{\mathrm{e}}
\newcommand{\ri}{\mathrm{i}}
\newcommand{\rd}{\mathrm{d}}
\RequirePackage{amssymb}
\begin{document}
\begin{center}
{\bf GEOMETRICAL FRAMEWORK OF QUANTIZATION PROBLEM   \\[1cm]}
{ M. Grigorescu \\[3cm]  }
\end{center}
\noindent
$\underline{~~~~~~~~~~~~~~~~~~~~~~~~~~~~~~~~~~~~~~~~~~~~~~~~~~~~~~~~
~~~~~~~~~~~~~~~~~~~~~~~~~~~~~~~~~~~~~~~~~~~}$ \\[.3cm]
The basic elements of the geometric approach to a 
consistent quantization formalism are summarized, 
with reference to the methods of the old quantum 
mechanics and the induced representations theory 
of Lie groups. A possible relationship between 
quantization and phase-space discretization is 
briefly discussed.  \\
$\underline{~~~~~~~~~~~~~~~~~~~~~~~~~~~~~~~~~~~~~~~~~~~~~~~~~~~~~~~~
~~~~~~~~~~~~~~~~~~~~~~~~~~~~~~~~~~~~~~~~~~~}$ 
\newpage
\section{Introduction}
The notion of quantization has appeared at the beginning of 
the last century in the theory of thermal  radiation, when 
M. Planck has formulated the hypothesis of energy 
quanta\footnote{It is important to remark that unlike the Nyquist 
formula for a linear electric circuit (RLC), the Planck distribution 
(accurately retrieved in the 2.7 K cosmic microwave background spectrum) contains 
no zero-point energy term, and the photon energy $\epsilon = {\rm h} \nu ={\rm  h} 
c / \lambda$ corresponds to a free relativistic particle with momentum 
$\vert {\bf p} \vert = {\rm h} / \lambda$ and Hamiltonian $c \vert {\bf p} \vert$ 
(arXiv0912.1998). Also,  for thermal radiation the wavelength $\lambda $ is 
the significant variable, as the Wien displacement law $\lambda_{max}T ={\rm  h}c/ 
4.965 k_B= 2.9$ mm$\cdot$K  describes the maximum of the spectral density 
(in vacuum) expressed as a function of $\lambda$.}: 
$\epsilon = {\rm h} \nu$, ${\rm h}=6.626 \times 10^{-34}$ J$\cdot$s \cite{mpl}. 
The existence of ${\rm h}$ was considered 
in statistical mechanics as evidence for a granular structure 
of the $2n$-dimensional phase-space, composed of elementary 
cells ("quantum states") of volume ${\rm h}^n$. For integrable 
systems with multiple periodic motions, such as the  hydrogen 
atom, this structure was provided by the quantization rules of 
the old quantum mechanics. Relativistic effects have also been 
included, as a  correction to Balmer's formula, due to the 
variation of mass with velocity, was introduced by Bohr 
\cite{bohr}, and the relativistic Kepler problem 
was quantized by Sommerfeld, applying integrality 
constraints to the action invariants \cite{somm}.  \\ \indent
In the algebraic (Dirac) approach to quantum mechanics, the 
observables are represented by elements of the set ${\cal F}(M)$ 
of smooth real functions over the classical (momentum) phase-space $(M, \omega)$, 
$M=T^*Q$, ($Q={\mathbb R}^3$), with 
$\omega$ the globally defined symplectic form.  Let $X_f$ 
be the vector field determined by $i_{X_f} \omega =  {\rd} f$, 
and ${\sf L}_X$ the Lie derivative with respect to $X$.  As 
${\cal F}(M)$ becomes a Lie algebra with respect to the 
Poisson bracket $\{ * , * \}$, $\{ f,g \} = \langle 
df , X_g \rangle =  \omega(X_f,X_g) = - {\sf L}_{X_f} g,~~f,g 
\in {\cal F}(M)$,   the full quantization of $M$ was defined 
as a ${\mathbb R}$-linear map $f \mapsto \hat{f}$ from ${\cal 
F}(M) $ to a set  ${\cal A}({\cal H})$ of symmetric operators 
on the Hilbert space ${\cal H}$, having the following properties 
\cite{am}: \\
1. the map $\hat{~}: {\cal F}(M) \mapsto {\cal A}({\cal H})$ 
is injective. \\
2. $[\hat{f}, \hat{g} ] = {\ri} \hbar \hat{~} \{ f,g \} $ 
$~;~$  $f,g \in {\cal F}(M)$. \\
3. $\hat{1} = I$, for $f=1$, constant on $M$, and $I$ the 
identity operator on ${\cal H}$. \\
4. $\hat{q}_k$, $\hat{p}_k$, $k=1,2,3$ act irreducibly on 
${\cal H}$. \\ \indent
It is presumed that once $\hat{~}$ and ${\cal H}$ are found, 
the quantum  dynamics with respect to the classical  time 
\cite{pfq, br} is given by the Schr\"odinger equation, and 
the scalar product in ${\cal H}$ has the statistical 
interpretation of probability amplitude.  \\ \indent
In classical nonrelativistic statistical mechanics, a Brownian 
particle can be described by a time-dependent distribution 
function ${\sf f} \ge 0$ defined on the phase-space $M= T^*
{\mathbb R}^3$, evolving according to the Fokker-Planck equation. 
Though,  at zero temperature  both classical and quantum 
distributions arise as  two different types of "functional 
coherent states" ${\sf f}_0$, ${\sf f}_\psi$  for the classical 
Liouville equation  \cite{cpw}. These are solutions associated 
with "action waves"  ${\sf n}^{[S]}$, respectively "quantum 
waves"  $\psi = \sqrt{{\sf n}} \exp( {\ri}  S / \hbar )$, 
expressed in terms of only two functions of coordinates and 
time: the localization probability density in the coordinate 
space ${\sf n}({\bf q},t)$, and the local "momentum potential" 
$S({\bf q},t)$.  Moreover, these classes are related, as the 
action distributions turn into quantum distributions (Wigner 
functions) when the configuration space  $Q={\mathbb R}^3$ is 
discretized. However, by contrast to the action distributions, 
 the Wigner distributions ${\sf f}_\psi$  remain  the same as 
functionals  of  $\psi$ during time evolution  only for 
polynomial potentials of degree at most 2. This limitation also 
appears in the canonical quantization, as the van Hove  
theorem  \cite{am, gs} states the incompatibility between the 
conditions 1,2,3,4. Thus, it is possible to fulfill the first 
three conditions, obtaining a "prequantization", but then the 
Heisenberg algebra ${\frak H} \equiv \{ q,p,1 \}$ is 
represented with infinite multiplicity. Also, if only the last 
three conditions are retained, then the map $\hat{~}$ should 
be restricted to some subalgebra ${\cal F}_P  \subset {\cal 
F}(M)$, containing ${\frak H}$. \\ \indent
This work\footnote{The next two sections are based on the notes of the seminar "Classical limit and quantization methods" given in 1989 at the Institute of Atomic Physics from Bucharest. The introductory section of this seminar, not included here, can be found in \cite{pfq}.}   presents, following  \cite{bk} as main reference, 
the geometrical framework in which the "action" and the 
"quantum" phase-space distributions are defined.  
The main concepts applied to the prequantization of Hamiltonian dynamical systems are recalled in Section 2. The reduction of the prequantum Hilbert 
space ${\cal H} \simeq {\rm L}^2(M, \omega^n)$, to the quantum 
Hilbert space  ${\cal H}_P \simeq {\rm L}^2(Q)$, is considered 
in Section 3. The transition from classical to quantum  
distributions  by phase-space discretization  is outlined  in 
Section 4.  Conclusions are summarized in Section 5.      
\section{The Prequantization }
\subsection{Equivalence Classes of Line Bundles }
Let $M$ be a $C^\infty$ differentiable manifold, separable and 
connected.  A line bundle on $M$ is a vector bundle  
\begin{center} 
\( \begin{array}{cc}
{\mathbb C} \mapsto & L \\
 &\downarrow  \pi\\  
& M     
\end{array} \). 
\\ 
\end{center} 
The projection map $\pi$ is smooth, and for any $ p \in M$, 
$L_p = \pi^{-1} (p)$ (the fiber in $p$) is a one-dimensional 
vector space over ${\mathbb C}$. \\ \indent
On $L$, as manifold, we can introduce local coordinates. Let 
${\cal U} = \{ U_i , i \in I \} $, be an open covering of $M$, 
and $s_i : U_i \mapsto L$ smooth non-vanishing sections, such 
that the map $\sigma_i : {\mathbb C} \times U_i \mapsto 
\pi^{-1} (U_i)$,  $\sigma_i(z,p) = z s_i(p)$, is a 
diffeomorphism. The set of pairs $\{(U_i, s_i), i \in I \}$ 
defines a local system for the bundle $L$. 
\\ \indent        
Let $\Gamma_L$ be the space of smooth sections $s: M 
\mapsto L$. For the local system $(U_i,s_i)$ any $s \in 
\Gamma_L$ can be written on $U_i$ as $s=\psi_i s_i$, where 
$\psi_i \in {\cal  F}_c (U_i)$ is a complex function on $U_i$. 
The collection $\{ \psi_i \}_{ i \in I}$ 
represents the local coordinates of $s$. \\ \indent
On $U_i \cap U_j$ the local system defines by the relation 
$s_i =c_{ij} s_j$ the transition functions $c_{ij} \in {\cal 
F}_c^*(U_i \cap U_j)$. These functions should satisfy the 
relationships 
\begin{equation}
c_{ij}=c_{ji}^{-1}, \qquad c_{ij}c_{jk}=c_{ik}
\end{equation}
on $U_i \cap U_j$, respectively on $U_i \cap U_j \cap U_k$. If 
expressed in the form $c_{ij} = \exp ({\ri} q_{ij} / \hbar )$, 
($\hbar = {\rm h}/ 2 \pi=1/2 \pi$, as we take ${\rm h}=1$), we can see 
that the new functions $q_{ij}$ provide a constant with 
integer values, denoted $a_{ijk}$, 
\begin{equation}
a_{ijk} = q_{ij}+q_{jk}-q_{ik} \in {\mathbb Z}
\end{equation}
on any intersection $U_i \cap U_j \cap U_k \ne \emptyset$. 
\\ \indent
Two line bundles  $L^1$ and $L^2$ on $M$ are equivalent if 
there exists a diffeomorphism $\tau : L^1 \mapsto L^2$ such 
that for any $ p \in M$, the map $\tau$ induces a linear 
isomorphism $L^1_p \mapsto L^2_p$. The set of equivalence 
classes of line bundles on $M$ is denoted ${\cal L}(M)$.  
\\ \indent
If $c^1_{ij}$, $c^2_{ij}$ are the transition functions for 
$L^1$, respectively $L^2$, then the two are equivalent iff 
there exists $\lambda_i = s^2_i/s^1_i$, $\lambda_i \in {\cal 
F}_c^* (U_i)$ (the set of nonvanishing complex functions on 
$U_i$), such that $c^2_{ij}=\lambda_i c^1_{ij} \lambda_j^{-1}$. 
Using this result it can be proved  \cite{bk} that there exists 
a one-to-one mapping $\kappa : {\cal L}(M) \mapsto H^2(M,
{\mathbb Z})$, which assigns to any element $\ell = [L] 
\in {\cal L} (M)$ the Cech cohomology class $[a] \in H^2(M, 
{\mathbb Z})$ \cite{fh} of the function $a_{ijk}$ associated 
with $L$. In particular, $L$ is called trivial if equivalent 
to ${\mathbb C} \times M$ ($\Gamma_L$ contains a nonvanishing 
global section).     
\subsection{Line Bundles with Connection} 
Let $\chi_c (M)$ be the Lie algebra of complex fields on $M$, 
and $L$ a line bundle on $M$. A connection in the line bundle 
$\pi :L \mapsto M$ is a linear map $\nabla :    \chi_c (M) 
\mapsto {\rm End} (\Gamma_L)$ such that 
\begin{equation}
\nabla_{\Phi \xi} = \Phi \nabla_\xi    \label{3} 
\end{equation}
\begin{equation}
\nabla_\xi (\Phi s) = ({\sf L}_\xi \Phi ) s + \Phi \nabla_\xi 
s , \qquad  \Phi \in {\cal F}_c(M),  \qquad  s \in \Gamma_L.
\end{equation} 
If $ \{ (U_i,s_i), i \in I \}$ is a local system for $L$, 
then $\nabla$ is completely specified by its action on the 
sections $\{s_i \}_{i \in I}$,
\begin{equation}
\nabla_\xi s_i = 2 \pi {\ri} \alpha_i  ( \xi ) s_i, \qquad   
\xi \in \chi_c(M), \qquad i \in I.
\end{equation} 
The condition (\ref{3}) implies $\alpha_i (\Phi \xi) = \Phi 
\alpha_i(\xi)$, such that the collection of functions $\{ 
\alpha_i(\xi)~, i \in I,~\xi \in \chi_c(M) \}$ defines a 
family of complex 1-forms $\{\alpha_i \}_{i \in I}$,  
$i_\xi \alpha_i \equiv \langle \alpha_i , \xi \rangle =  
\alpha_i  ( \xi )$,
associated to the connection $\nabla$. On $U_i \cap U_j$ we get 
$$\alpha_i = \alpha_j + {\rd} q_{ij}$$
and conversly, any family of 1-forms having this property 
specifies uniquely a connection $\nabla$. Such a family 
arises by the pull-back of an unique ${\mathbb C}^*$ - 
invariant 1-form $\alpha \in \Omega^1(L^*)$, called connection 
form. Here  $\Omega^k(L^*) $ denotes the set of $k$-forms on 
the manifold $L^* = \{ \cup_{p \in M} L^*_p ~;~ L^*_p 
= L_p - \{ 0 \} \}$. The form $\alpha$ is globally defined on 
$L^*$, and $s_i^* \alpha = \alpha_i$, $ i \in I$.   \\ \indent
If $(L^1, \alpha^1)$, $(L^2, \alpha^2)$, are line bundles on 
$M$ with connection forms $\alpha^1$, $\alpha^2$, then there 
exists a diffeomorphism $\tau: L^1 \mapsto L^2$ such that 
$\tau$ induces a linear isomorphism 
$$L^1_p \mapsto L^2_{\pi (\tau(L^1_p))}$$ 
and $\tau^* \alpha^2 = \alpha^1$. \\ \indent
One should note that any equivalence $(L, \alpha) \mapsto 
(L, \alpha^1)$ is specified by a function $\Phi \in {\cal 
F}_c^*(M)$, such that
$$
\tau_\Phi^* \alpha^1 = \alpha^1 - \frac{1}{2 \pi {\ri}} 
\frac{ {\rd} \tilde{\Phi}}{\tilde{\Phi}} = \alpha,  \qquad 
\tilde{\Phi} = \pi^*( \Phi)
$$
and $\tau_\Phi (x) = \Phi(\pi x) x $,  $ x \in L$. In 
particular, $ \tau_\Phi: (L, \alpha) \mapsto (L, \alpha)$ 
is an equivalence iff $\Phi$ is a complex constant on $M$. 
\\ \indent 
The family of 1-forms $\{\alpha_i \}_{i \in I}$, associated 
to the connection $\nabla$ determines an unique complex 2-form 
$\omega$ on $M$, such that 
\begin{equation}
{\rd} \omega =0, \qquad  \omega \vert_{U_i} = {\rd} \alpha_i, 
\qquad \pi^* \omega = {\rd} \alpha.
\end{equation}
Because 
$$
[\nabla_\xi, \nabla_\eta ] - \nabla_{[ \xi, \eta]} = 2 \pi 
{\ri} \omega( \xi, \eta),  \qquad   \xi, \eta \in \chi_c(M)
$$ 
$\omega$ is called the curvature form of the connection 
$\nabla$. If real and nondegenerate, $\omega$ provides a 
symplectic structure on $M$.         
\subsection{Line Bundles with Connection and Hermitian 
Structure} 
A Hermitian structure on $L$ is a function $(*,*): L 
\times L \mapsto {\mathbb C}$ having the properties: \\
$i)$ $(*,*)$ induces a structure of 1-dimensional Hilbert 
space on $L_p$, for all $ p \in M$. \\
$ii)$ $\vert * \vert^2$ is a positive function on $L^*$, 
$\vert x \vert^2  \equiv (x,x)$, $x \in L^*$. \\ \indent
Let $\gamma$ be a smooth curve on $M$. The covariant 
derivative of the section $s \in \Gamma_L$ along $\gamma$ 
is defined by
\begin{equation}
\frac{ Ds}{Dt} = \nabla_{\dot{\gamma}(t)} s.
\end{equation}    
For any smooth curve $\gamma$ on $M$, $\{ \gamma_t, t 
\in (a,b) \}$, the covariant derivative defines a linear 
isomorphism $\tau_{t',t}: L_{\gamma_t} \mapsto L_{\gamma_{t'}}$, 
called parallel transport by 
\begin{equation}
\frac{ Ds}{Dt} \vert_{\gamma_t }= \frac{{\rd}}{{\rd}  t'}  
\tau_{t,t'} s (\gamma_{t'}) \vert_{t'=t}.
\end{equation}    
A section $r= \psi s_i$  is autoparallel along $\gamma$  if 
$\nabla_{\dot{\gamma}} r =0$, or
$$ \psi(\gamma_t) = {\re}^{ - 2 \pi {\ri} \int_0^t \langle 
\alpha_i , \dot{\gamma}_t \rangle } \psi (\gamma_0). $$
If $\gamma = \partial \Sigma  \subset M$ is closed, 
contractible on $\Sigma$ to a point, then 
$${\cal Q}_\gamma =  {\re}^{ - 2 \pi {\ri} \int_\Sigma  
\omega }$$
is the scalar function of parallel transport.  Applications 
to autoparallel sections for the constrained quantum dynamics 
are presented  in \cite{cev}. \\ \indent
The Hermitian form $(*,*)$ is called $\nabla$-invariant if 
the parallel transport leaves invariant the inner product on 
fiber,
\begin{equation}
\frac{{\rd}}{{\rd} t} ( \tau_{t,t'} s^1_{(\gamma_{t'})}, 
\tau_{t,t'} s^2_{ (\gamma_{t'})}) \vert_{t'=t} = 0
\end{equation}    
or
\begin{equation}
{\sf L}_\xi ( s^1,s^2 )=  (\nabla_\xi s^1,s^2) +(s^1, 
\nabla_\xi s^2) .
\end{equation}    
When $s^1=s^2=s_i$ this reduces to
\begin{equation}
{\rd} \ln \vert s_i \vert^2 = 2 \pi {\ri} (\alpha_i - 
\bar{\alpha}_i) 
\end{equation}    
where $\bar{\alpha}_i$ is the complex conjugate of $\alpha_i$. 
Thus, $\alpha_i - \bar{\alpha}_i$ is a real 1-form, exact on 
$U_i$,  the curvature form 
$$
\omega \vert_{U_i} = {\rd} \alpha_i = {\rd} \bar{\alpha}_i
$$
is real, and ${\cal Q}_\gamma \in {\rm U}(1)$. Let 
$[ \omega ]_{dR} \in H^2_{dR}(M,{\mathbb R})$ be the de 
Rahm cohomology class of $\omega$. In general, the isomorphism 
between $H^2_{dR}(M,{\mathbb R})$ and $H^2(M,{\mathbb R})$ 
associates to a real, closed two-form $\omega$ on $M$, 
expressed locally as 
$$
\omega \vert_{U_i} = {\rd} \alpha_i, \qquad  \alpha_i = 
\alpha_j + {\rd} f_{ij}, \qquad  f_{ij}:M \mapsto {\mathbb R}     
$$ 
the class $[\omega] \equiv  [a^\omega]  \in H^2(M,{\mathbb R})$, 
where $a^\omega_{ijk} = f_{ij}+f_{jk}-f_{ik}$ is a real constant 
on $U_i \cap U_j \cap U_k$. However, if $\omega$ is the 
curvature of a connection $\nabla$ on a line bundle $L$ with 
$\nabla$-invariant Hermitian structure, then $a^\omega_{ijk}$ 
is an integer\footnote{ If $\omega_f$ and $\omega_f^*$ are ${\rm SO}(3,1)$-invariant, dual (electromagnetic) 2-forms on $ {\mathbb R}^{3,1}$, then  $[\omega_f]_{dR}=0$ (the first group of Maxwell equations) and $[\omega_f^*]_{dR}/e $ is integral (electric charge quantization, arXiv0912.1998).  }, and $\omega$ specifies an integral cohomology 
class in  $H^2(M,{\mathbb R})$. \\ \indent
Conversly, the problem is to what extent a closed, real 2-form 
$\omega$, satisfying an integrality condition, determines a 
Hermitian line bundle with connection on $M$. If $\omega$ is 
integral, then in general $a^\omega_{ijk}$ are not integers, 
but we can find real constants $x_{ij}=-x_{ji}$ on $U_i \cap 
U_j \ne \emptyset$, such that 
$$
z_{ijk}= a_{ijk} + x_{ij} +x_{jk}-x_{ik} 
$$
are integers on   $U_i \cap U_j \cap U_k \ne \emptyset$. This 
result allows one to define a line bundle $L$ on $M$ with the 
transition functions 
$$
c_{ij} = \exp (2 \pi {{\ri}} q_{ij}), \qquad q_{ij} = 
f_{ij} + x_{ij}
$$
on $U_i \cap U_j \ne \emptyset$. Because
$$
\alpha_i = \alpha_j + {\rd}  f_{ij}= \alpha_j + {\rd} q_{ij} 
= \alpha_j + \frac{1}{2 \pi {\ri}} \frac{{\rd} c_{ij}}{c_{ij}}
$$
with $\alpha_i, \alpha_j$ real, there exists on $L$ a 
connection $\nabla$ defined by the family of 1-forms 
$\{ \alpha_i \}_{ i \in I }$, and a $\nabla$-invariant 
Hermitian structure. \\ \indent
In this formulation, the 1-forms $\alpha_i$ are defined by 
$\omega$ up to a total differential ${\rd} \Phi_i$. If 
$\alpha_i'=\alpha_i + {\rd} \Phi_i$, then $f'_{ij} =f_{ij}+ 
\Phi_i-\Phi_j$, and
$$
c'_{ij}= \lambda_i c_{ij} \lambda_j^{-1}, \qquad \lambda_i 
= {\re}^{2 \pi {\ri} \Phi_i}
$$
define a Hermitian line bundle with connection $(L', \nabla')$,
equivalent to $(L, \nabla)$. In specifying this equivalence 
class there is still an arbitrary due to the way of choosing 
the constants $x_{ij}$. Thus, the integrality condition allows 
one to replace $x_{ij}$ by new real constants $x'_{ij}=x_{ij} 
+y_{ij}$, where $y_{ij}+y_{jk}+y_{ki} \in {\mathbb Z}$,
and $y_{ij}=-y_{ji}$. The line bundle $L'$, specified by the 
transition functions 
$$
c'_{ij}= \exp{ 2 \pi {\ri} (f_{ij}+x'_{ij})}= {\re}^{2 \pi 
{\ri} y_{ij}} c_{ij} 
$$ 
is equivalent to $L$ only if $y_{ij}$ has the form $y_{ij} = 
c_i -c_j$. Because   
$y_{ij}+y_{jk}+y_{ki} \ne 0$, $y_{ij}$ does not specify a 
cocycle in $C^1(M,{\mathbb R})$, but in the exponential it 
determines a cocycle in $C^1(M,{\rm U}(1))$. The bundles $L'$ 
and $L$ are equivalent only if this cocycle is coboundary, 
such that the set of equivalence classes of the Hermitian line 
bundles whose connection has the same curvature form $\omega$ 
is parameterized by $H^1(M,{\rm U}(1))$. This set of 
equivalence classes is denoted by ${\cal L}_c (M, \omega)$, 
and the result presented above states the isomorphism 
${\cal L}_c (M, \omega) \simeq H^1(M, {\rm U}(1)) 
\simeq H^2(M, {\mathbb Z})$. \\ \indent
Let $\epsilon : H^2(M, {\mathbb Z}) \mapsto H^2(M,{\mathbb R})$ 
be the homomorphism induced by the injection  $\epsilon : 
{\mathbb Z} \mapsto {\mathbb R}$, $\kappa : {\cal L}(M) 
\mapsto H^2(M, {\mathbb Z})$ the bijection introduced in 
subsection 2.1, and $\sigma: {\cal L}_c \mapsto {\cal L}$ 
the mapping given by $\sigma [(L, \alpha)] = [L]$. In this 
case, the Weil integrality condition states that if $\omega$ 
is any real, closed 2-form on $M$, then: \\
$i)$ ${\cal L}_c (M, \omega) \ne \{ \emptyset \}$ iff 
$[\omega] \in H^2(M,{\mathbb R})$ is integral. \\
$ii)$ $\sigma {\cal L}_c(M, \omega) = \{ [L] \in {\cal L} 
~;~ \epsilon \kappa [L]= [\omega] \}$. \\ 
Applications to the calculus of the energy levels for the 
multidimensionl Kepler problem are presented  in 
\cite{ml1,ml2,ml3}.
\subsection{The BWS Condition }
Let $(N, \omega)$ be a reducible presymplectic manifold, and 
$(M', \omega')$, with $M'=N/K$, the reduced space. Here $K$ 
is a smooth distribution on $N$, with the tangent space 
$$ T_mK = \{ x \in T_mN  ~;~ i_x \omega_m =0 \}.$$
{\bf Proposition 1.} A sufficient condition to obtain a 
quantizable reduction $(M', \omega')$ of $(N, \omega)$ is      
\begin{equation}
\oint_\gamma \theta \in {\mathbb Z}   \label{bws}
\end{equation}
where $\theta$ is a global 1-form such that $\omega = {\rd} 
\theta$, and $\gamma$ is any closed curve contained in a leaf 
of $K$. If $N$ is simply connected, then  (\ref{bws}) is also 
necessary \cite{nw}.  \\
For the proof we take a contractible covering ${\cal U}= \{U_i,
 i\in I \}$ of $M'$, such that for any $ i \in I$ there exists 
a section $\Sigma_i$ in $K$ over $U_i$ and a diffeomorphism 
$\rho_i : U_i \mapsto \Sigma_i$. If $m_1,m_2 \in U_i \cap U_j$ 
are two points joined by the curve $c$, then $\rho_i (c)$ is a 
curve in $\Sigma_i$, and $\rho_j (c)$ is a curve in $\Sigma_j$. 
Moreover, $\rho_i(m_1)$ and $\rho_j(m_1)$ can be joined by a 
curve $\gamma_1$ in the leaf of $K$ through $m_1$, respectively 
$\rho_i(m_2)$ and $\rho_j(m_2)$ can be joined by a curve 
$\gamma_2$ in the leaf of $K$ through $m_2$. Let $S$ be the 
surface bounded by 
$\rho_i(c), \rho_j(c), \gamma_1, \gamma_2$, so that $\pi(S)=c$.
Because $S \in \ker ( \omega)$,   
$$
\int_S \omega = 0 = \oint_{\partial S} \theta = 
\int_{\rho_i(m_1)}^{\rho_i(m_2)} \theta     
- \int_{\rho_j(m_1)}^{\rho_j(m_2)} \theta   
$$
$$
+ \int_{\rho_i(m_2)}^{\rho_j(m_2)} \theta - 
\int_{\rho_i(m_1)}^{\rho_j(m_1)} \theta = 
f_{ji}(m_2) - f_{ji}(m_1)+ \int_{m_1}^{m_2} (\rho_i^* \theta 
- \rho_j^* \theta)    
$$ 
which yields 
$$
\rho_i^* \theta - \rho_j^* \theta \equiv \theta_i - \theta_j 
= {\rd} f_{ij} . 
$$
The 1-forms $\theta_i =\rho_i^* \theta$, $\theta_j =\rho_j^* 
\theta$ on $U_i \cap U_j  \ne \{ \emptyset  \}$ are related 
to the symplectic form $\omega' = {\rd} \theta_i = {\rd} 
\theta_j  $. The functions $f_{ij}= - f_{ji}$ on $U_i 
\cap U_j $ can be defined by integration along an 
arbitrary curve contained in the leaf of $K$ over $m$,  
$$f_{ij}(m) = \int_{\rho_i(m)}^{\rho_j(m)} \theta .$$
Thus, $f_{ij}+f_{jk}+f_{ik} \in {\mathbb Z}$ as an integral 
(\ref{bws}) of the 1-form $\theta$ along a closed curve in 
the leaf of $K$ through $m$, proving that the class 
$[\omega'] \in H^2(M',{\mathbb R})$ is integral.  \\ \indent
When $N = h^{-1} (E) \subset M$ is the constant energy surface 
of a classical system on $(M, \omega)$ with Hamiltonian $h$, 
then (\ref{bws}) is similar to the Bohr-Wilson-Sommerfeld 
(BWS) condition from the old quantum mechanics.  

\subsection{The Prequantum Hilbert Space and Operators Related 
to Observables}
Let $(M, \omega)$ be a quantizable classical phase-space, in 
the sense that 
$[\omega] \in H^2(M,{\mathbb R})$ is integral. In this case, 
on $M$ we can define a Hermitian line bundle with connection 
$(L, \alpha)$. The natural volume element on $M$ is  
$\epsilon_\omega = \omega^n$, and for $\omega = 
\sum_{k=1}^n {\rd} q_k \wedge {\rd} p_k$,
$$
\epsilon_\omega = {\rd} q_1 \wedge ... \wedge {\rd} 
q_n \wedge {\rd} p_1 \wedge ... \wedge {\rd} p_n. 
$$
If $(*,*)$ denotes the $\nabla$-invariant Hermitian form on 
$L$, the prequantum Hilbert space ${\cal H}$ is defined as 
the space of all sections $s \in \Gamma_L(M)$ for which
$$
\int_M \epsilon_\omega~ (s,s)  
$$
exists and is finite. The inner product in ${\cal H}$ is 
\begin{equation}
< s_1, s_2 > \equiv \int_M \epsilon_\omega ~(s_1, s_2),   
\qquad  s_1,s_2 \in {\cal H}. \label{ipr}
\end{equation}
Let ${\frak e}(L)$ be the Lie algebra of the ${\mathbb 
C}^*$-invariant, real fields on $L^*$. By the existence 
of the connection form $\alpha$ and the projection $\pi_* 
: TL \mapsto TM$, there exists also a linear isomorphism 
\begin{equation}
{\frak e}(L) \mapsto {\cal F}_c(M) \times \chi (M)
\end{equation}
which associates to $ \eta \in {\frak e}(L)$ a function 
$\Phi \in {\cal F}_c(M)$ and a vector $\xi \in \chi(M)$ 
such that
\begin{equation}
\pi^* \Phi = - \langle \alpha , \eta \rangle, \qquad  
\xi = \pi_* \eta.
\end{equation}
Conversly, any function $\Phi \in {\cal F}_c(M) $ specifies 
an unique field $\eta_\Phi \in {\frak e}(L)$, $\eta_\Phi 
\in \ker (\pi_*)$ by the relation
$$
\langle \alpha, \eta_\Phi \rangle = - \pi^* \Phi
$$
and for any $ \xi \in \chi(M)$ there exists an unique 
field $\hat{\xi} \in {\frak e}(L)$, $\hat{\xi}_x \in 
\ker (\alpha_x)$, $ x \in L^*$   \cite{bk}
$$
\pi_* \hat{\xi} = \xi .
$$ 
{\bf Proposition 2.} ${\frak e}(L)$ is parameterized by 
${\cal F}_c(M) \times \chi(M)$, such that for any $ \Phi 
\in {\cal F}_c(M)$ and $ \xi \in \chi(M)$,
\begin{eqnarray}
\eta_{(\Phi, \xi)} & = & \eta_\Phi + \hat{\xi} \in 
{\frak e}(L)
\\
\lbrack \eta_{ ( \Phi_1 , \xi_1 ) }, \eta_{ ( \Phi_2 , 
\xi_2) } \rbrack &
= & \eta_{(\xi_1 \Phi_2 - \xi_2 \Phi_1 + \omega (\xi_1,\xi_2), 
[\xi_1,\xi_2])}       
\end{eqnarray}
{\bf Proof.} Let us denote by $\eta_{ ( \Phi , \xi ) }$ the 
commutator $[\eta_{ ( \Phi_1 , \xi_1 ) }, \eta_{ ( \Phi_2 , 
\xi_2) }]$. Then
\begin{equation}
\pi_* \lbrack \eta_{ ( \Phi_1 , \xi_1 ) }, \eta_{ ( \Phi_2 , 
\xi_2) } \rbrack = \lbrack \xi_1, \xi_2 \rbrack \equiv \xi
\end{equation}
and
\begin{equation}
\langle \alpha, \lbrack \eta_{ ( \Phi_1 , \xi_1 ) }, 
\eta_{ ( \Phi_2 , \xi_2) } \rbrack 
\rangle \equiv - \pi^* \Phi.  \label{19}
\end{equation}
With the identity 
\begin{equation}
\langle \alpha, \lbrack \eta_1, \eta_2 \rbrack \rangle = 
{\sf L}_{\eta_1} \langle \alpha, \eta_2 \rangle  - 
{\sf L}_{\eta_2} \langle \alpha, \eta_1 \rangle - {\rd} 
\alpha ( \eta_1, \eta_2)
\end{equation}
(\ref{19}) becomes 
\begin{equation}
\Phi = {\sf L}_{\xi_1} \Phi_2   - {\sf L}_{\xi_2} \Phi_1 + 
\omega ( \xi_1, \xi_2).  \square
\end{equation} \indent
The elements of the algebra ${\frak e}(L)$ act on functions on 
$L^*$, but we can also find a representation of ${\frak e}(L)$ 
in the space of the sections $\Gamma_L$. Thus, we can define 
an  ${\frak e}(L)$-isomorphism $\tilde{~}:\Gamma_L(M) \mapsto 
{\cal F}(L^*)$ associating to any section $s \in \Gamma_L(M)$ 
a function $\tilde{s} \in {\cal F}(L^*)$, 
$$\tilde{s} (x) = \frac{s( \pi x)}{x}, \qquad  x \in L^* .$$
{\bf Proposition 3.} If $ (\Phi, \xi) \in {\cal F}_c(M) \times 
\chi(M)$ then $ \eta_{(\Phi, \xi)} \tilde{s} = \tilde{t}$, 
where
$$ t = (\nabla_\xi + 2 \pi {\ri} \Phi ) s  \equiv \hat{\eta} 
s .$$
With respect to a local system  $\{ (U_i,s_i), i \in I \}$ on 
$L$, the elements of $\Gamma_L$ are represented by functions 
$\psi_i :U_i \mapsto {\mathbb C}$, provided by 
$$s\vert_{U_i} = \psi_i s_i, \qquad  s \in \Gamma_L. $$
For this local trivialization, the operator 
$\hat{\eta}_{(\Phi, \xi)} = \nabla_\xi + 2 \pi {\ri} \Phi $
determines an operator  $\hat{\eta}_{i (\Phi, \xi)}$ on 
${\cal F}_c(U_i)$    
$$\hat{\eta}_{i(\Phi, \xi)} = {\sf L}_\xi + 2 \pi {\ri} 
(\langle \alpha_i , \xi \rangle + \Phi ). $$       
The tangent fields to $L$ which preserve the connection and 
the Hermitian structure form a subalgebra in ${\frak e}(L)$, 
denoted ${\frak e}(L, \alpha)$. It can be proved \cite{ bk} 
that if $\eta_{(\Phi, \xi)} \in {\frak e}(L)$, then
\begin{equation}
{\sf L}_{\eta_{(\Phi, \xi)}} \vert * \vert^2 =0~~{\rm 
iff}~~\Phi \in {\cal F}(M)
\end{equation}
and
\begin{equation}
{\sf L}_{\eta_{(\Phi, \xi)}} \alpha = \pi^*(i_\xi \omega  
- {\rd} \Phi).
\end{equation}
This shows that
$$ \eta_{(\Phi, \xi)} \in  {\frak e}(L, \alpha)~~ 
{\rm iff}~~\Phi \in {\cal F}(M)~~{\rm and}~~ i_\xi 
\omega ={\rd} \Phi.$$ 
Therefore, the mapping 
$$\delta : {\cal F}(M) \mapsto {\frak e}(L, \alpha) 
~;~ \delta( \Phi) = \eta_{(\Phi, \xi_\Phi)} 
$$
(or ${\cal F}(M) \mapsto {\frak end}(\Gamma_L)$), called 
map of prequantiztion, is an isomorphism of Lie algebras. 
\\ \indent
These results indicate that we can obtain a representation 
of the Lie algebra of the observables, ${\cal F}(M)$, in 
the prequantum Hilbert space ${\cal H}$. In this 
representation each function $\Phi$ has an associated 
operator  
$$\hat{\eta}_{(\Phi, \xi_\Phi)} = \nabla_{\xi_\Phi} + 
2 \pi {\ri} \Phi $$
on $\Gamma_L$, or on the space of the local representatives 
$\psi_i$ of the sections,
$$\hat{\eta}_{i(\Phi, \xi_\Phi)} = {\sf L}_{\xi_\Phi} + 
2 \pi {\ri} (\langle \alpha_i , \xi_\Phi \rangle + \Phi ). $$
Thus, defining the local operator associated to  the 
observable $f$ as 
\begin{equation}
\hat{f} = \frac{1}{2 \pi {\ri}} \hat{\eta}_{(f, \xi_f)} 
\equiv \frac{1}{2 \pi {\ri}} {\sf L}_{\xi_f} + \langle 
\alpha_i , \xi_f \rangle + f
\end{equation}
one obtains a map which satisfies the conditions 1,2,3 
stated in the Section 1, discussed in detail in \cite{pfq}. 
 In particular, if $M = {\mathbb R}^2$, $\alpha_i = - p 
{\rd} q$, then $\xi_p = \partial_q$,
$\xi_q = - \partial_p$, and $\hat{p} = - {\ri} \hbar 
\partial_q$, $\hat{q} = q+  {\ri}  \hbar \partial_p$. 

\subsection{The Prequantization of Classical Dynamical Systems}
The classical dynamical systems on the phase-space $(M,\omega)$
are subgroups of ${\cal D}(M)$, the group of diffeomorphisms 
on $M$. The symplectic diffeomorphisms form a subgroup 
denoted ${\cal D}(M, \omega)$, of diffeomorphisms which 
act by canonical transformations,
$${\cal D}(M, \omega) = \{ \rho \in {\cal D}(M) ~;~ 
\rho^* \omega = \omega \}.$$
This subgroup contains  ${\rm Ham}(M, \omega)$, the subgroup 
of Hamiltonian diffeomorphisms, and 
if $M$ is simply connected, or if $TM = [TM,TM]$, then
${\cal D}(M, \omega)= {\rm Ham}(M, \omega)$. \\ \indent
The phase-space $(M, \omega)$ has also an associated set of 
equivalence classes of Hermitian line bundles with connection, 
${\cal L}_c (M, \omega)$. The group ${\cal D}(M, \omega)$ acts 
on ${\cal L}_c (M, \omega)$, but prequantum representations in 
a class $\ell \in {\cal L}_c (M, \omega)$, can be obtained only 
for the elements of the stability group ${\cal D}_\ell (M, 
\omega)$ of  $\ell$ with respect to the action of ${\cal D} 
(M, \omega)$,
$${\cal D}_\ell (M, \omega) = \{ \rho \in { \cal D}
 (M, \omega) ~;~ \rho^*_{\cal L} \ell = \ell, 
\ell \in {\cal L}_c (M, \omega) \}.$$ 
Thus, if $[L]=\ell$ and $\rho \in {\cal D}_\ell (M, \omega)$, 
then $\rho^*_{\cal L} L$ and $L$ are equivalent, and there 
exists an equivalence  of line bundles with connection 
$\epsilon: \rho^*_{\cal L} L \mapsto L$, uniquely specified  
up to a phase factor. \\ \indent
In general, if $G$ is a group acting on $(M, \omega)$ by 
canonical transformations, there are operators $\hat{g} 
: {\cal H} \mapsto {\cal H}$ which define a projective 
representation of $G$ in ${\cal H}$, such that for 
$ g_1,g_2 \in G$,
\begin{equation}
 \hat{g_1g_2}= \tau_{12} \hat{g}_1 \hat{g}_2, 
\qquad \vert \tau_{12}\vert=1.
\end{equation}
In the prequantum Hilbert space ${\cal H} \subset \Gamma_L$, 
$[L]=\ell$, the operator associated to $\rho \in {\cal D}_\ell 
(M, \omega)$  is defined up to a phase factor by the equality
\begin{equation}
(\hat{\rho}^{-1} s)_{(p)} = \epsilon (\rho^* s_{(p)}).
\end{equation} 
If $\rho_h(t) \in {\rm Ham}(M,\omega)$ is generated by the 
Hamiltonian $h$, then $\rho^* L=L$ and the operator 
$\hat{\rho}_t : \Gamma_L \mapsto \Gamma_L$ will be defined by
\begin{equation}
\tilde{\rho}_\eta(t) \circ \hat{\rho}_t^{-1} s = s \circ 
\rho_h(t)   \label{hrho}
\end{equation}       
where $\tilde{\rho}_\eta (t)$ is the group of one-parameter 
diffeomorphisms on $L^*$ determined by $\eta_{(h, X_h)} \in 
{\frak e}(L, \alpha)$.  \\
{\bf Theorem 1.}  $\eta_{(h, X_h)} \in {\frak e}(L, \alpha)$ 
is globally integrable on $L^*$ iff $X_h$ is globally 
integrable on $M$, and the diagram 
\begin{center} 
\( \begin{array}{cccc}
\tilde{\rho}_\eta(t) : & L^* & \mapsto & L^* \\
& \pi\downarrow &  & \pi\downarrow \\  
~~~~~\rho_t: & M & \mapsto & M  \\   
\end{array} \) 
\\ 
\end{center} 
commutes.  \\ \indent
To obtain explicitly the operator $\hat{\rho}_t$, we can write 
(\ref{hrho}) in local coordinates. Let $\{ (U_i,s_i), i\in I 
\}$ be a local system, and  $\sigma$  the diffeomorphism 
$$\sigma : {\mathbb C} \times U_i \mapsto \pi^{-1} (U_i), 
\qquad  \sigma(z,p) = z s_i(p).$$
The functions $\tilde{s}(x) \equiv s(\pi x)/x$ on $L^*$ 
associated to the sections $s \in \Gamma_L(U_i)$ are 
represented locally by functions $\tilde{s}^\flat$ on 
${\mathbb C} \times U_i$, 
$$ \tilde{s}^\flat (z,p) = \tilde{s}(z s_i(p)) = \frac{1}{z} 
\tilde{s} (s_i(p))= \frac{1}{z} \psi (p).$$
Also, the connection form $\alpha$ and the field $\eta_{(\Phi, 
\xi)}$ have the local expressions 
\begin{equation}
\alpha^\flat = \alpha_i + \frac{1}{2 \pi {\ri}} \frac{{\rd} 
z}{z},  \qquad \alpha_i = s_i^* \alpha
\end{equation}
\begin{equation}
\eta^\flat =  \xi^\flat - 2 \pi {\ri}   \Phi   z \partial_z  
\end{equation}  
where $\xi^\flat= \xi - 2 \pi {\ri}  (  \langle \alpha_i, \xi 
\rangle   z \partial_z 
-  \langle \bar{ \alpha}_i, \xi \rangle  \bar{z} 
\partial_{\bar{z}}) \in \ker(\alpha^\flat)$.
The flow of $\eta^\flat$ determines the time-evolution of 
the functions $\tilde{s}^\flat_i$ by the equation
\begin{equation}
\frac{{\rd} \tilde{s}^\flat_i}{{\rd} t} = \eta^\flat 
\tilde{s}^\flat_i.
\end{equation}
This provides the dependence on time of the coordinate $z$ and 
of the point $p \in U_i$ in terms of the local expression, denoted 
$\tilde{\rho}^\flat_{\eta^\flat} (t)$,  
\begin{equation}
\tilde{\rho}^\flat_{\eta^\flat} (t) (z_0,p_0) = (z_0 
{\re}^{ - 2 \pi {\ri} \int_0^t dt' (  
\langle \alpha_i, \xi \rangle+  \Phi)}, \rho_t (p_0))   
\label{31}   
\end{equation}
of the flow $\tilde{\rho}_\eta (t)$. \\ \indent
The operator $\hat{\rho}(t)$ defines an operator $\hat{U}_t$ 
acting on the complex functions $\psi (p)=s(p)/s_i(p)$, 
representing the sections $s\in \Gamma_L(U_i)$, by 
\begin{equation}
\hat{U}_t \psi = \frac{\hat{\rho}_t s}{s_i}.
\end{equation}
Explicitly, this is obtained from (\ref{hrho}) in local form,
\begin{equation}
\tilde{\rho}^\flat_\eta (t) (\hat{U}^{-1}_t \psi ,p) = (\psi 
(\rho_t p), \rho_t (p))
\end{equation}
where the action of  $\tilde{\rho}^\flat_\eta$ is given by 
(\ref{31}),
\begin{equation}
\tilde{\rho}^\flat_\eta (t) (\hat{U}^{-1}_t \psi ,p) = 
({\re}^{ - 2 \pi {\ri} \int_0^t {\rd} t' ( \langle \alpha_i, 
\xi \rangle+ \Phi)} (\hat{U}^{-1}_t \psi)_p, \rho_t (p)).
\end{equation}
The result 
\begin{equation}
 (\hat{U}^{-1}_t \psi)_p = {\re}^{2 \pi {\ri} \int_0^t {\rd} 
t' ( \langle \alpha_i, \xi \rangle+ \Phi)} \psi ( \rho_t (p))
\label{35}
\end{equation}
agrees with the expression derived in the previous subsection 
for the local operator 
$$\hat{\Phi} =  \frac{1}{2 \pi {\ri}} {\sf L}_{\xi_\Phi} + 
\langle \alpha_i , \xi_\Phi \rangle + \Phi$$
because
\begin{equation}
{\ri} \hbar \frac{{\rd}}{{\rd} t}( \hat{U}_t \psi) = 
\hat{\Phi} \hat{U}_t   \psi.
\end{equation}
\subsection{Applications to Elementary Systems }
Let $G$ be a simply connected Lie group with the Lie algebra  
${\frak g}$, and  ${\frak g}$$^*$ the dual of  ${\frak g}$.
For $f \in {\frak g}^*$ one can define on $G$ a right ($R_g$) 
- invariant 1-form $\theta_f$, and a closed 2-form $\omega_f 
= {\rd} \theta_f$,
$$
\omega_f(x,y) \vert_e = \langle f , [x,y]  \rangle, \qquad 
x,y \in  {\frak g}. $$
The distribution $K_f$ determined by the kernel of $\omega_f$ 
on $G$ has as tangent space at the identity $e$
$$T_e K_f = \{ x \in  {\frak g}  ~;~  \langle f , 
[x,y] \rangle =0 ~{\rm for~ any~} y \in  {\frak g} \} $$ 
namely the algebra ${\frak g}_f$ of the stability group $G_f$ 
of $f$ with respect to the coadjoint action of $G$. Thus, the 
leaf of $K_f$ through $e$ is the connected component $(G_f )_0$ 
of $G_f$, and if closed,   $M' =G/K_f  \simeq  G/(G_f )_0 $ is 
covering space for the orbit $M_f=G/G_f$ of $f$ in  
${\frak g}^*$. \\ 
{\bf Theorem 2.} 
Let $(M', \omega')$  be the reduced phase-space
associated to the reducible presymplectic manifold $(G, 
\omega_f)$.  Then  $(M', \omega')$ is quantizable iff $f$ can 
be integrated to a character for $(G_f)_0$.  \\
{\bf Proof}. Let us assume first that  
$$ \oint_\gamma \theta_f \in {\mathbb Z} $$     
(the BWS condition) with $\gamma \subset (G_f)_0$. Thus, one 
can define
\begin{equation}
\chi_f (h) = {\re}^{2 \pi {\ri} \int_e^h \theta_f }   \label{t1.1} 
\end{equation}
where the integral can be taken along any curve in $(G_f )_0$, 
which joins $e$ to $h$. Because 
\begin{equation}
\chi_f (h_1h_2) = {\re}^{2 \pi {\ri} \int_e^{h_1h_2} 
\theta_f }= {\re}^{2 \pi {\ri} \int_e^{h_2} \theta_f +
2 \pi {\ri} \int_{h_2}^{h_1h_2} \theta_f} ~~ 
\end{equation}
independently of the integration path, from the BWS condition, 
while  
$$\int_{h_2}^{h_1h_2} \theta_f = \int_e^{h_1} \theta_f $$
from the $R_g$-invariance of $\theta_f$, one obtains 
\begin{equation}
\chi_f (h_1h_2) = \chi_f(h_1) \chi_f(h_2)
\end{equation}
such that $\chi_f$ is a character for $(G_f)_0$. If 
$h = {\re}^{tx}$, with $x \in {\frak g}_f$, then
$$ \chi_f ({\re}^{tx}) = {\re}^{2 \pi {\ri} \langle f, x 
\rangle t}$$
from the $R_g$-invariance, such that $f$ appears as an 
infinitesimal character in the sense that
\begin{eqnarray}
\frac{{\rd} }{{\rd} t}  \chi_f({\re}^{tx}) \vert_{t=0} 
& = & 2 \pi {\ri} \langle f, x \rangle, 
\qquad  x \in {\frak g}_f  \label{t1.2} \\
\langle f, [x,y] \rangle &= &0, \qquad  x,y \in {\frak g}_f.
\end{eqnarray}
Conversly, the condition to integrate the infinitesimal 
character $f$ to a character of $(G_f)_0$ independently 
of the path, applied to (\ref{t1.1}) derived from 
(\ref{t1.2}), leads to the BWS condition $\square$. 
 \\ \indent
Let us consider
\begin{equation} 
G={\rm SU}(2) = \{ \left[  \begin{array}{cc} z_0 & z_1 
\\ - \bar{z}_1 & \bar{z}_0  \end{array} \right] 
~;~ \vert z_0\vert^2+ \vert z_1\vert^2 =1 \} 
\simeq  {\mathbb S}^3 \subset {\mathbb C}^2.  
\end{equation}
The algebra  ${\frak g} \equiv T_e G$ of $G$ consists of matrices 
$$ x_a =- \frac{{\ri}}{2}\left[  \begin{array}{cc} a_1 & a_2- 
{\ri} a_3 \\ a_2+ {\ri} a_3 & -a_1  \end{array} \right]  = 
\sum_{i=1}^3 a_i  E_i  $$
with $(a_1,a_2,a_3) \equiv {\bf a}  \in {\mathbb R}^3$ and 
$[E_i, E_j ] = \epsilon_{ijk} E_k$. \\ \indent
The right (left) - invariant vector fields $Y_a$ ($Z_a$) 
extending $x_a \in {\frak g} $ are\footnote{If $x_a, x_b \in {\frak g}$ and 
$[x_a, x_b]=x_c$, then $[Z_a, Z_b]=Z_c$,  $[Y_a, Y_b]=-Y_c$.} 
\begin{equation}
Y_a = - \frac{{\ri}}{2} \lbrack (a_1 z_0 -(a_2- 
{\ri} a_3) \bar{z}_1) \partial_{z_0}+ 
(a_1 z_1 +(a_2- {\ri} a_3) \bar{z}_0) \partial_{z_1} 
\rbrack + c.c.
\end{equation}
\begin{equation}
Z_a = - \frac{{\ri}}{2} \lbrack (a_1 z_0 +(a_2+ {\ri} a_3) z_1)
\partial_{z_0}+ (-a_1 z_1 +(a_2- {\ri} a_3) z_0) \partial_{z_1}
\rbrack + c.c.
\end{equation}
and the right-invariant 1-form $\theta_f$ associated to $ f 
\in {\frak g}^* \simeq {\frak g}  $   is
\begin{equation}
\theta_f = {\ri} \lbrack (f_1 \bar{z}_0 -(f_2+ {\ri} f_3) 
z_1) {\rd} z_0+ (f_1 \bar{z}_1 +(f_2+ {\ri} f_3) z_0) {\rd} 
z_1 \rbrack + c.c.
\end{equation}
where $c.c.$ is the complex conjugate of the previous term. In 
particular, for ${\bf f} \equiv (f_1,f_2,f_3)=(-l,0,0)$ we get
$$
(G_f)_0= G_f = \{ \left[  \begin{array}{cc} {\re}^{{\ri} t} 
& 0 \\ 0 & {\re}^{- {\ri} t}  \end{array} \right] ~;~ 
t \in {\mathbb R} \} \subset G  
$$ 
and
\begin{equation}
\theta_f = {\ri} l  \sum_{k=0}^1 (z_k {\rd} \bar{z}_k - 
\bar{z}_k {\rd} z_k), \qquad \omega_f = 2 {\ri}  l 
\sum_{k=0}^1 {\rd} z_k \wedge {\rd} \bar{z}_k.
\end{equation}
Each $h_t\in G_f$ is generated by  $x_a \in {\frak g}$ with 
${\bf a} =(-2,0,0)$,  
$$x_a =\left[  \begin{array}{cc} {\ri} & 0 \\ 0 & - {\ri}  
\end{array} \right], \qquad h_t={\re}^{tx_a} = 
\left[  \begin{array}{cc} {\re}^{{\ri} t} & 0 \\ 0 & 
{\re}^{- {\ri} t}  \end{array} \right] $$ 
and as
$$ \langle \theta_f ,  Y_a \rangle =  \langle f , x_a  
\rangle = {\bf f} \cdot {\bf a} = - a_1 l =2l $$
it determines a character 
$$\chi_f (h_t) = {\re}^{ 2 \pi {\ri}  \langle \theta_f , Y_a 
\rangle t} = {\re}^{ 4 \pi {\ri} l t}.$$
This character allows one to define a line bundle $L'$ on 
$M'=G/G_f$ by factorizing the trivial bundle $G \times 
{\mathbb C}$ with respect to the equivalence relation "$\sim$",
$$ (g,z) \sim (hg,\chi_f(h) z) $$
where $g \in G$, $h \in G_f$ and $z \in {\mathbb C}$. The 
sections in $\Gamma_{L'}$ are represented by functions 
$\psi:G \mapsto {\mathbb C}$ (sections in $G\times 
{\mathbb C}$) which satisfy the global relation 
\begin{equation}
\psi (h g) = \chi_f (h) \psi(g)  \label{46}
\end{equation}
or locally
\begin{equation}
Y_a \psi(g) = 2 \pi {\ri} {\bf a} \cdot {\bf f} \psi (g).
\end{equation} 
Thus, the sections of $\Gamma_{L'}$ are represented in the 
coordinates $(z_i,\bar{z}_i)$ by functions $\psi: {\mathbb 
S}^3 \mapsto {\mathbb C}$ which satisfy
\begin{equation}
\sum_{k=0}^1 (z_k \partial_{z_k} - \bar{z}_k 
\partial_{\bar{z}_k} ) \psi(z, \bar{z}) = 4 \pi 
l \psi(z, \bar{z}).
\end{equation}
The equivalence relation "$\sim$" is well defined, and $M'$ 
is quantizable if
$$\chi_f (h_{2 \pi})= {\re}^{8 \pi^2 i l} =1, \qquad  
4 \pi l  \in {\mathbb Z} $$ 
namely $l=n \hbar /2$ (here $\hbar  = 1/2\pi$), with $n \in 
{\mathbb Z}$.  A physical application  to the intrinsic angular
 momentum (spin)  is presented in \cite{sn}.  \\ \indent
The points of the phase-space $M'$ correspond to equivalence 
classes in $G$ defined by 
$$ [g] = \{ hg ~;~ h \in G_{\bf f}, g \in G \}.$$
Let ${\rm pr}:G \mapsto M'$ be the projection ${\rm pr}(g) 
=[g]$, $ g \in G$. A canonical action  $$g_1[g] =[gg_1^{-1}]$$ 
of $G$ on $M'$ can be defined by the projection on $M'$ of the 
action to the right of $G$ on $G$ (because the equivalence 
necessary for projection is obtained by the action to the 
left), and $M'$ becomes a homogeneous phase-space for $G$. 
Locally, the action of $G$ on $M'$ arises by the projection 
of the left-invariant fields, $-Z_a$, on $TM'$, ${\rm pr}_{*} 
(-Z_a) = X_a$, and because the algebra  ${\frak g}$ is 
semisimple, there exists a lift $\lambda$ of this action such 
that the diagram 
\begin{center} 
\( \begin{array}{ccc}
 0  \mapsto & {\cal F} (M') \mapsto       {\frak ham}(M')       
&  \mapsto  0 \\ 
~~~~~~~         & {\lambda}  \nwarrow   \uparrow ~~& \\ 
~~~~~~~         & ~~~~                            
{\frak g}~~ &  
\end{array} \) 
\\ 
\end{center} 
commutes. Explicitly, for any  $x_a \in {\frak g}$ one can 
find $h_a:M' \mapsto {\mathbb C}$, ${\rm pr}^* h_a = \langle  
\theta_{\bf f} , Z_a \rangle$, representing the Hamiltonian 
of the field $X_a$,
$$  i_{X_a} \omega ' = {\rd} h_a    .$$
To get the time-evolution of the sections from the line bundle 
$L'$, associated with the dynamical system generated on $M'$ 
by the Hamiltonian $h_a$, we project on $L'$ the trajectory 
in $G \times {\mathbb C}$ of the dynamical system generated  
by the Hamiltonian $\langle \theta_f, Z_a \rangle$. This 
trajectory is given by (\ref{35}) in which 
$$\alpha_i = \theta_ f, \qquad \xi=-Z_a, \qquad \Phi= \langle 
\theta_f , Z_a \rangle $$               
\begin{equation}
(\hat{U}_{g_t} \psi )(g) = \psi (gg_t),  \qquad g \in G
\end{equation}
while the projection on $\Gamma_{L'}$ requires $\psi$ 
constrained by (\ref{46}),
\begin{equation}
\psi (h g) = \chi_f (h) \psi(g), \qquad g \in G, 
\qquad h \in G_f .
\end{equation}
The result indicates that the prequantization of the phase-space 
$(M', \omega')$ is equivalent to the derivation of the 
representations of the group $G$ induced by the character 
$\chi_f$ of the subgroup $G_f$ \cite{gwm}. In general these 
representations are not irreducible (do not provide a 
quantization for $(M', \omega')$), but imposing the condition 
as $\psi$ to be holomorphic, we obtain irreducible 
representations. Thus, the holomorphy condition, by 
introducing a complex polarization, represents a way of 
restricting the prequantum Hilbert space.  \\ \indent
The technique of the induced representations was successfully 
applied to quantize the relativistic free particle or the liquid 
drop. In both cases the classical configuration space is the 
orbit of a group $H$ in a linear space $V$, and the 
quantization consists of finding induced representations for 
the semidirect product $G = H \times V$. In the first case 
$H={\rm O}(3,1)$ is the Lorentz group, $V= {\mathbb R}^{3,1}$ 
is the Minkowski space, and $G$ is the Poincar\'e group, while 
in the second case  $H={\rm SL}(3, {\mathbb R})$, $V={\mathbb 
R}^4$, and $G={\rm CM}(3)$ \cite{rowe, ri, ml5}.  The case of 
a massive free particle in the anti-de Sitter spacetime is 
considered in \cite{sdb}.
\section{Elements of Quantization}
\subsection{Complex Polarizations}    
A complex polarization of the $2n$-dimensional manifold $(M, 
\omega)$ is a complex distribution $P$ having the following 
properties: \\
$i$) for any $ m \in M$, $P_m \subset T_m^cM$ is a complex 
Lagrangian subspace. \\   
$ii$) $D_m = P_m \cap \bar{P}_m \cap T_m M$ has a constant 
dimension. \\
$iii$) $P$ is integrable, in the sense that for any $ m \in M$ 
there exists a collection of functions $\{ z_k \in {\cal 
F}_c(M), k=1,n \}$, such that $\{ \bar{X}_{z_k}, k=1,n \}$ 
generate $P_m$. \\ \indent
Let us introduce the notation
$$ \chi_c (U,P) = \{ X \in \chi_c (U)  ~;~  X_m 
\in P_m,~~ m \in U \subset M \}  $$    
$$ {\cal F}_c (U,P) = \{ f \in {\cal F}_c (U)  ~;~  
\bar{X} f =0,~~ X \in \chi_c(U,P) \}$$
$$ =\{ f \in {\cal F}_c (U)  ~;~  \bar{X}_f \in 
\chi_c(U,P) \}  $$    
$$ {\cal F}_c (U,P,1) = \{ f \in {\cal F}_c (U)  ~;~  
\{f,g \} \in {\cal F}_c (U \cap V,P), 
~ V \subset M, ~ g \in {\cal F}_c (V,P) \} .$$
The set $ {\cal F}_c (U,P,1)$ consists of functions having the 
property that generate flows which preserve the polarization,
\begin{equation}
{\sf L}_{X_f} \bar{P} \subset \bar{P}  \Leftrightarrow f \in 
{\cal F}_c (U,P,1).
\end{equation}
When $f\in {\cal F}_c (U,P,1) $ is real, the flow of 
$X_f$ preserves both $\bar{P}$ and $\omega$. The 
polarization $P$ is called admisible if on a neighborhood of 
any $ m \in M$, there exists  a symplectic potential $\beta$, 
which is adapted to $P$ in the sense that 
$$i_{\bar{X}} \beta =0, \qquad   X \in \chi_c(M,P).$$
The polarization $P$ is of K\"ahler type if $P_m \cap 
\bar{P}_m = \{ 0 \}$ and $T_mM = P_m+ \bar{P}_m$. In the 
K\"ahler case any $ X \in T_mM$ can be written as $X=Z+
\bar{Z}$, with $Z \in P_m$, and $T_mM$ carries a complex 
structure, 
$$ J_m : T_mM \mapsto T_mM, \qquad J_mX= {\ri} Z- {\ri} 
\bar{Z}$$
compatible with the symplectic form $\omega$  in the sense that
$ \omega(JX,JY)= \omega(X,Y)$. \\ \indent
Let $(M,\omega,J)$ be a K\"ahler manifold (Appendix \cite{gs}),
and $\{ z_k , k=1,n \}$ local complex coordinates such that 
\begin{equation}
J \partial_{z_k} = {\ri} \partial_{z_k}, \qquad J 
\partial_{\bar{z}_k} =-  {\ri} \partial_{\bar{z}_k}.
\end{equation}
On $M$ can be introduced two polarizations: the holomorphic 
polarization $P$ generated at any point by the vectors $\{ 
\partial_{z_k}, k=1,n \}$, and the antiholomorphic polarization
 $\bar{P}$ generated by $\{ \partial_{\bar{z}_k}, k=1,n \}$. 
Thus, for any $ U \subset M$,
$$ {\cal F}_c(U,P) = \{ f:U \mapsto {\mathbb C}  ~;~  
f= {\rm holomorphic} \}.$$ 

\subsection{Phase-Space Quantization for K\"ahler Polarizations}
Let $(M,\omega)$ be a symplectic manifold,  $\{ q_k, p_k, 
k=1,n \}$ the local canonical coordinates for $\omega$ 
($\omega = \sum_{k=1}^n {\rd} q_k \wedge {\rd} p_k$), and  
$P$  a K\"ahler polarization on $M$ locally generated by the 
vectors
$$\{ \partial_{z_k} ~;~ z_k= (q_k - {\ri}  p_k) / 
\sqrt{2},~ k=1,n\}.$$          
The complex potential adapted to $P$ is 
\begin{equation}
\beta = {\ri}  \sum_{k=1}^n \bar{z}_k {\rd}  z_k 
\end{equation}
and $\omega = {\rd} \beta$. If $(M, \omega)$ is quantizable, 
then there exists a Hermitian line bundle with connection $(L, 
\alpha)$ on $M$, with  the sections space $\Gamma_L(M)$ and 
associated prequantum Hilbert space ${\cal H}$. We can further 
define  the space of polarized sections
\begin{equation}
\Gamma_L (M,P) = \{ s \in \Gamma_L(M)  ~;~ 
\nabla_{\bar{X}} s=0,~~  X \in \chi_c(M,P) \}
\end{equation}    
and the quantum Hilbert space $ {\cal H}_P = {\cal H} \cap 
\Gamma_L(M,P)$.
The space $\Gamma_L(M,P)$ is well defined because the local 
integrability condition  for the sections $s \in \Gamma_L(M,P)$ is satisfied. Thus, if 
$$\nabla_{\bar{X}} s = \nabla_{\bar{Y}} s = 0, \qquad X,Y 
\in \chi_c(M,P)$$
then 
\begin{equation}
\nabla_{[\bar{X}, \bar{Y}]} s = [ \nabla_{\bar{X}}, 
\nabla_{\bar{Y}}] s - 2 \pi {\ri} \omega (\bar{X}, \bar{Y}) s 
=0 
\end{equation}
because $\omega (\bar{X}, \bar{Y})=0$. \\ \indent
The Hilbert space ${\cal H}_P$ is not invariant to the action 
of any operator associated with a classical observable, and 
one should specify which classical observables provide 
operators on ${\cal H}_P$. If $f \in {\cal F}(M)$,  then
$$ \hat{f} = \frac{1}{2 \pi {\ri}} \nabla_{X_f} +  f $$
and the condition $\hat{f} {\cal H}_P \subset {\cal H}_P$ 
yields
\begin{equation}
\nabla_{\bar{X}} \hat{f} s =0, \qquad  X \in \chi_c (M,P),
  \qquad  s \in {\cal H}_P. \label{56}
\end{equation}
However, because
$$ 2 \pi {\ri} \nabla_{\bar{X}} \hat{f} s = \nabla_{\bar{X}} 
( \nabla_{X_f} + 2 \pi {\ri} f) s =
([ \nabla_{\bar{X}}, \nabla_{X_f}] + 2 \pi {\ri} 
{\sf L}_{\bar{X}} f )s 
$$
$$
=(\nabla_{[\bar{X},X_f]}+ 2 \pi {\ri} \omega(\bar{X},X_f) + 
2 \pi {\ri} {\sf L}_{\bar{X}} f )s = \nabla_{[\bar{X},X_f]} 
s $$
the condition (\ref{56}) is equivalent to ${\sf L}_{X_f} 
\bar{P} \subset \bar{P}$.  Thus, an observable $f$ determines 
an operator on ${\cal H}_P$ only   if $f \in {\cal F}_c (M,P,1)$.
\\ \indent
Let $s$ be a section  of  $\Gamma_L(M,P)$ for which  $s^* 
\alpha = \beta$, and $r$ the unit section in $\Gamma_L (U)$ 
such that $r^* \alpha = \sum_{k=1}^n q_k  {\rd} p_k$. Then, 
$s={\re}^{ \varphi} r$, with $\varphi$ specified (up to an 
additive constant $\varphi_0$) by
$$ \nabla_X s = 2 \pi {\ri} \langle \beta , X \rangle s = 
\nabla_X ({\re}^{ \varphi} r)= ({\sf L}_X \varphi) s + 2 
\pi {\ri} \langle r^* \alpha , X \rangle s  $$
with
$$ \langle \beta , X \rangle = \langle r^* \alpha , X 
\rangle + \frac{1}{2 \pi {\ri}} \langle {\rd} \varphi , 
X \rangle, \qquad  X \in \chi (M) $$       
such that
$$ {\rd} \varphi = 2 \pi {\ri} (\beta - r^* \alpha) = 2 
\pi {\ri} \sum_{k=1}^n [ \frac{{\ri} }{2} (q_k+ {\ri} p_k) 
{\rd} (q_k- {\ri} p_k) - q_k {\rd} p_k] $$ 
$$= - \frac{\pi}{2} \sum_{k=1}^n {\rd} (q_k^2+p_k^2 +2 {\ri}  
q_kp_k) = - \pi \sum_{k=1}^n {\rd} (\vert z_k \vert^2 + {\ri}  
q_k p_k). $$   
Considering $\varphi_0=0$, we get $\varphi = - \pi 
\sum_{k=1}^n (\vert z_k \vert^2 + {\ri} p_k q_k )$ and
$$(s,s) = {\re}^{- 2 \pi \sum_{k=1}^n \vert z_k \vert^2}.$$
Thus, with respect to the local system specified by $s$, the 
elements of the space ${\cal H}_P$ are sections of the form 
$\{ \psi_p s_p, ~p \in U \subset M \}$, where $\psi_p$ are 
holomorphic functions of $\{ z_k,~ k=1,n \}$, and the inner 
product (\ref{ipr}) is given by
$$ < \psi_1, \psi_2 > \sim \int \epsilon_\omega 
\bar{ \psi}_1(z) \psi_2(z) {\re}^{-2 \pi \sum_{k=1}^n 
\vert z_k \vert^2 }. $$  
This Hilbert space coincides with the representation introduced
in 1928 by V. Fock, to study the states of the harmonic 
oscillator. Though, its domain of applicability remains limited
because the only observables quantizable in ${\cal H}_P$ are 
polynomials in coordinates and momenta of degree at most 2. 
\\ \indent
For the harmonic oscillator the classical Hamiltonian is 
$h = 2 \pi \nu \bar{z} z $, and
$$ X_h = 2 \pi {\ri} \nu ( z \partial_z - \bar{z} 
\partial_{\bar{z}}).$$
The operator $\hat{h}$ in ${\cal H}$ associated with $h$,
$$ \hat{h} = \frac{1}{2 \pi {\ri}}  {\sf L}_{X_h} +  \langle 
\beta, X_h \rangle +   h = \nu  (z \partial_z - \bar{z} 
\partial_{\bar{z}}) $$
becomes $\hat{h}_P = \nu z \partial_z $ when restricted to 
${\cal H}_P$. Its eigenvalues are $n \nu$,  $n \in {\mathbb 
Z}$, showing that this approach yields the same result, 
physically incomplete, as the old quantum mechanics.   
To obtain the missing term $\nu /2$ the bundle of polarized 
sections  should be extended by a line bundle  of half-forms 
providing the measure for the inner product in ${\cal H}_P$, 
as indicated on examples in \cite{ml4}.   
\subsection{Real Polarizations and Asymptotic Solutions}
A real polarization of the symplectic manifold $(M, \omega)$ is
 a foliation of $M$ by Lagrangian (maximal isotropic) 
submanifolds. If $M=T^*Q$, and $\omega= \sum_{k=1}^n  
{\rd }q_k \wedge {\rd} p_k$ is the canonical 2-form, then the 
vertical foliation $P$ is a real polarization, and the leaves 
of $P$ are the surfaces $q_k=$constant, $k=1,n$. \\ \indent
Let $P$ be a real polarization of the symplectic manifold $(M, 
\omega)$. Then, on a neighborhood of any point $ m $ of $ M$ 
one can find  canonical coordinates $(x,y) \equiv 
(x_k,y_k)_{k=1,n} $  such that the leaves of $P$ coincide 
locally with the surfaces  $x=$ constant or $y=$ constant. 
The canonical coordinates having this property are called 
"adapted to $P$". 
\\ \indent
Let $\Lambda \subset M$ be a Lagrangian submanifold and 
$U \subset M$ containing $\Lambda$ such that $\omega \vert_U = {\rd} \theta$. 
Because $\omega \vert_\Lambda =0$ then also ${\rd} \theta 
\vert_\Lambda =0$, and locally there exists a function $\wp$ 
on $\Lambda$, called "local phase function", $\wp: \Lambda 
\mapsto {\mathbb R}$, such that $\theta \vert_\Lambda = -{ 
\rd} \wp$. \\ \indent
If $M = T^*Q$, then $\Lambda$ is transversal to the vertical 
polarization $P$ if the restriction to $\Lambda$ of the 
projection    
\begin{center} 
\( \begin{array}{cc}
\Lambda \subset & M \\
 &\pi \downarrow \\  
& Q     
\end{array} \) 
\\ 
\end{center} 
is a diffeomorphism. In this case $S \in {\cal F}(W)$,
 $\pi ( \Lambda) = W \subset Q$,  $\pi^* S = \wp$, 
is called "generating function of the first kind" of $\Lambda$. 
Moreover, $\Lambda \cap T^* Q$ determines a 1-form on 
$W$ with the local  coordinates 
$$(p,q) \equiv (\frac{\partial S}{\partial q}, q) .$$
Thus, a foliation of the phase-space $M=T^* Q$ by Lagrangian 
submanifolds corresponds to a family of generating functions 
$S (q,y)$, $y \equiv \{ y_k,~k=1,n \}$, parameterized by the 
variables  $y$. This type of foliation appears naturally in 
classical mechanics by the Hamilton-Jacobi equation, 
$$ h ( \partial_q S, q) = {\rm constant}  $$     
which represents the condition $ h\vert_{\Lambda_S} = {\rm 
constant} $
for the Lagrangian submanifold $\Lambda_S$ of $T^*Q$ generated 
by $S$.  \\
{\bf Proposition 4.} Let $\Lambda \subset M$ be a connected 
Lagrangian submanifold of the phase-space $(M, \omega)$ and 
$h \in {\cal F}(M)$. Then $h$ is a constant on $\Lambda$ iff 
$X_h \in \chi (M)$ is tangent to $\Lambda$.  \\ \indent
If we denote $x \equiv \{ x_k = \partial S / \partial y_k, 
k=1,n \}$, then $(x,y)$ is a local coordinate system on 
$T^*Q$ adapted to the polarization $\Lambda_S$ determined 
by $S(q,y)$. In this system $h$ is a function only of $y$, 
and the equations of motion are 
$$ \dot{y} =0, \qquad \dot{x} ={\rm constant}.$$
In particular, when  $Q \simeq {\mathbb R}^n$ and $\Lambda_S 
\subset {\mathbb T}^n \equiv {\mathbb R}^n / {\mathbb Z}^n$ 
is part of an invariant torus, then $y$-$x$ are the 
"action-angle" coordinates used to express the BWS 
conditions.  \\ \indent
To quantize a classical system described by the Hamiltonian 
$h$ it is convenient to find a Hilbert space 
${\cal H}_{\Lambda_S}$ associated to the polarization 
determined by the solution $S$ of the Hamilton-Jacobi 
equation. Because in the case $\Lambda_S \subset 
{\mathbb T}^n$ the BWS  conditions  provide constraints on 
the phase-space reduction (to a point) and the stationary 
states,  it is natural to select the sections from 
$\Gamma_L(M,\Lambda_S)$ by  $\nabla_X r=0$, where $X$ is 
tangent to $\Lambda_S$ and $r \in \Gamma_L(M)$. Let $s$ be a 
section in $\Gamma_L(M)$ such that 
$$ s^* \alpha \vert_{U_i} = - {\rd} S  $$
and $r= \psi s$ an arbitrary element in 
$\Gamma_L(M,\Lambda_S)$. The equation
\begin{equation}
 \nabla_X r = ({\sf L}_X \psi) s - 2 \pi {\ri} 
\langle {\rd} S, X \rangle \psi s = ({\sf L}_X \psi) s - 
2 \pi {\ri} ({\sf L}_X S) \psi s =0 
\end{equation}
has the solution $\ln \psi - 2 \pi {\ri} S = f(y)$, where $f$ 
is an arbitrary function of $y$, or
\begin{equation}
\psi (q,y) = a (y) {\re}^{2 \pi {\ri} S(q,y)}.
\end{equation}
The sections from $\Gamma_L(M,\Lambda_S)$ can be transferred 
to the space $\Gamma_L(M,P)$, where $P$ is the vertical 
polarization associated to the Schr\"odinger representation. 
The function obtained \cite{nw}
\begin{equation} 
\Psi (q) = A(q) {\re}^{2 \pi {\ri}  S(q)}
\end{equation}
can be interpreted as asymptotic solution of the Schr\"odinger 
equation in the WKB \cite{vor} approximation.  \\ \indent
As $h$ may contain any potential, in general $S$ is 
multiple-valued, and  $\Psi$ should be defined by a sum 
$\sum_\mu  A_\mu  \exp ( 2 \pi {\ri} S_\mu )$ over different 
branches. Presuming  that $A_\mu$  can have the branches  
$\pm \vert A_\mu \vert$, but $\Psi$ remains single-valued, 
one obtains corrected BWS quantum conditions \cite{jbk}, which 
yield for the harmonic oscillator the exact  energy levels  
$\nu (n + 1/2)$, $n=0,1,2,...~$.      
\section{Quantization and Discretization}
The geometric elements presented in the previous sections 
also appear  in the formalism of statistical mechanics. Let 
${\sf f} \ge 0$ be the distribution function \cite{as} of a 
classical system composed of N identical, non-interacting 
particles, defined on the one-particle phase-space  
$(M, \omega)$, normalized by 
\begin{equation}
\int_M \epsilon_\omega  {\sf f}({\bf q},{\bf p},t) = {\rm N}, 
\qquad {\rm N} \geq 1.  \label{nc}
\end{equation} 
For a one-particle Hamilton function $h: M \mapsto 
{\mathbb R}$, at zero temperature and without friction, 
${\sf f}({\bf q},{\bf p},t)$ evolves according to the 
transport equation 
\begin{equation}
\partial_t {\sf f} + {\sf L}_{X_h} {\sf f} =0. \label{te}
\end{equation}
Let us consider $M=T^*{\mathbb R}^3$, $\omega =  \sum_{i=1}^3 
{\rd} q_i \wedge {\rd} p_i$,   $h({\bf q},{\bf p})= 
{\bf p}^2/2m + V({\bf q})$ ,  and
 \begin{equation}
{\sf f} ({\bf q},{\bf p},t)= \frac{1}{(2 \pi)^3} \int {\rd}^3 
k ~{\re}^{- {\ri} {\bf k} \cdot {\bf p}}~~ \tilde{\sf f} ({\bf 
q},{\bf k},t)    \label{ift}
\end{equation}
where $\tilde{\sf f} ({\bf q},{\bf k},t)$ is the Fourier 
transform of ${\sf f} ({\bf q},{\bf p},t)$. In this case, a 
particular class of exact solutions for (\ref{te})  are the 
"action distributions" ${\sf f}_0({\bf q} ,{\bf p},t)$, 
provided by  
\begin{equation}
\tilde{\sf f}_0({\bf q} ,{\bf k},t) = {\sf n}({\bf q} ,t) 
{\re}^{{\ri} {\bf k} \cdot {\bf \partial_q} S({\bf q} ,t)}   
\label{f0}
\end{equation}
where ${\sf n}({\bf q}, t) \geq 0$ (the particle density) and  
$S({\bf q}, t)$   satisfy the continuity, respectively the 
Hamilton-Jacobi equations \cite{cpw}. \\ \indent 
The partial derivative ${\bf k} \cdot {\bf \partial_q} 
S({\bf q} ,t)$ in (\ref{f0}) is the limit of 
\begin{equation} 
\frac{k}{\ell} [ S({\bf q}+ \frac{\ell}{2k}  {\bf k},t)-S({\bf 
q} - \frac{\ell}{2k} {\bf k},t)]  \label{fd} 
\end{equation}
 $k= \vert {\bf k} \vert$, 
when $\ell \rightarrow 0$. If a new parameter $\sigma = 
\ell / k$ is introduced, then    
\begin{equation}
\tilde{\sf f}_0({\bf q} ,{\bf k},t) = \lim_{\sigma \rightarrow 
0} \tilde{\sf f}_\psi({\bf q} ,{\bf k},t)
\end{equation}
where
\begin{equation}
\tilde{\sf f}_\psi ({\bf q} ,{\bf k},t) \equiv \psi^*({\bf q} 
- \frac{\sigma {\bf k}}{2},t) \psi ({\bf q} + \frac{\sigma 
{\bf k}}{2},t) \label{lim0}
\end{equation}
and $\psi = \sqrt{{\sf n}} \exp( i S / \sigma )$.  However, 
when $k \rightarrow 0$ 
$$  S({\bf q} \pm  \frac{\sigma_0}{2}  {\bf k} ,t) = S({\bf 
q},t) \pm  \frac{\sigma_0}{2}  {\bf k} \cdot {\bf \partial_q} 
S({\bf q} ,t) + \frac{\sigma_0^2}{8} ({\bf k} \cdot {\bf 
\partial_q})^2  S({\bf q} ,t) \pm ...
$$
and if the terms containing $ (\sigma_0 k)^m$, $m \geq 3$ are 
neglected, then 
$${\bf k} \cdot {\bf \partial_q} S({\bf q} ,t) =  
\frac{1}{\sigma_0} [ S({\bf q}+  \frac{\sigma_0 }{2} 
{\bf k},t)-S({\bf q} - \frac{ \sigma_0}{2} {\bf k},t)]  $$
for any dimensional constant  $\sigma_0$. Thus, within a 
suitable domain of ${\bf k}$, we may also consider in 
(\ref{lim0}) $\sigma$ as a finite constant, (e.g. $\sigma 
= \hbar$), such that ${\sf f}_\psi$ defined by (\ref{ift}), 
\begin{equation}
{\sf f}_\psi({\bf q},{\bf p},t)= \frac{1}{(2 \pi)^3} \int 
d^3k ~{\re}^{- {\ri} {\bf k} \cdot {\bf p}} ~~\tilde{\sf 
f}_\psi ({\bf q} ,{\bf k},t)  
\end{equation}
is the Wigner transform \cite{gut} of $\psi({\bf q} ,t)$. In 
this case, the normalization condition (\ref{nc}) takes the 
form 
\begin{equation}
\int {\rd}^3q {\rd}^3p ~~{\sf f}_\psi ({\bf q},{\bf p},t) = 
\int d^3q ~~ \vert \psi ({\bf q} ,t) \vert^2 =  \langle \psi 
\vert \psi \rangle  = {\rm N}
\end{equation} 
and the phase-space overlap between two distributions 
${\sf f}_{\psi_1}$, 
${\sf f}_{\psi_2}$, (resembling the inner product (\ref{ipr})), 
is \cite{rpw}
\begin{equation}
<{\sf f}_{\psi_1} {\sf f}_{\psi_2} > \equiv \int 
{\rd}^3q {\rd}^3p~~ {\sf f}_{\psi_1} {\sf f}_{\psi_2} = 
\frac{ \vert \langle \psi_1 \vert \psi_2 \rangle \vert^2}{(2 
\pi \sigma)^3} 
\end{equation} 
where 
\begin{equation}
\langle \psi_1 \vert \psi_2 \rangle \equiv \int {\rd}^3q~~ 
\psi_1^*({\bf q} ,t) \psi_2({\bf q} ,t)  . 
\end{equation}
Worth noting, within this framework can be defined overlaps 
$<{\sf f}_{0_1} {\sf f}_{0_2} >$ between  "action distributions", 
or mixed overlaps $<{\sf f}_{\psi} {\sf f}_0 >$, while
${\sf f}$ can be a sum ${\sf f} = {\sf f}_0 + {\sf f}_\psi$.
 \\ \indent
Usually, derivatives are replaced by finite differences such 
as (\ref{fd}) in numerical  or lattice \cite{wen} calculations,
as a result of discretization.  Variations in the  length 
unit (measure), with respect to a connection form provided by the 
electromagnetic potentials, have also been introduced by H. Weyl  
\cite{weyl, cartan}. Presuming that canonical and mechanical 
momentum coincide, let $\{ {\bf a}_1, {\bf a}_2, {\bf a}_3 \}$ 
be the (covariant) fundamental vectors of a Bravais lattice 
${\cal B}_P$ in the momentum space,  $\{ {\bf a}^1, {\bf a}^2, 
{\bf a}^3 \}$ the (contravariant) fundamental vectors of the 
reciprocal lattice ${\cal B}_P^*$,  
$$ {\bf a}^l= 2 \pi \epsilon_{lmn} \frac{ {\bf a}_m 
\times {\bf a}_n }{ \Omega_P}, \qquad \Omega_P =   
{\bf a}_1 \cdot ( {\bf a}_2 \times {\bf a}_3) $$
and  $\{ {\bf c}^i =  \hbar {\bf a}^i, i=1,2,3 \}$ the 
fundamental vectors of a Bravais lattice ${\cal B}_Q$ in 
the coordinate space.  Thus, if $\tilde{\sf f} ({\bf q},
{\bf k},t)$ has significant values only when ${\bf k}$ 
relates two nodes of ${\cal B}_P^*$ and $\sigma {\bf k}$ in 
(\ref{lim0}) relates two nodes of ${\cal B}_Q$, then $\sigma 
= \hbar$. 
 \\ \indent
These considerations indicate that a natural relationship 
between the classical distribution function ${\sf f}_0$ and 
the quantum WKB wave function  arises by phase-space 
discretization in elementary cells of volume $\Omega_P \Omega_Q = {\rm h}^3$, 
$\Omega_Q = {\bf c}^1 \cdot ({\bf c}^2 \times {\bf c}^3)$, using for the 
canonically conjugate variables a direct lattice and its reciprocal. It can also be shown  
\cite{cpw} that the Wigner function ${\sf f}_\psi$  is an 
exact solution of  (\ref{te}) with  $\psi$ an  exact solution 
of the Schr\"odinger equation, only if the potential $V$ is a 
polynomial of degree at most 2.     
\section{Summary and Conclusions}
The integrality conditions of the old quantum mechanics, as 
well as the correspondence between observables and operators, 
sought in algebraic quantization, may receive a geometrical 
interpretation in the theory of complex line bundles with 
connection and Hermitian structure. The basic elements of 
this "prequantization" theory have been presented in 
Section 2. However, the meaning of the complex line over 
a (physical or reduced) phase-space $(M, \omega)$ is not 
obvious. The distribution function ${\sf f} \ge 0$ used 
in classical statistical mechanics is integrable over $M$, 
but real, while the quantum wave functions $\psi$ are complex, 
but $\bar{ \psi} \psi$  is integrable only over the 
configuration space.  To retrieve this property, the 
prequantum Hilbert space can be reduced, as indicated in 
Section 3,  by selecting a polarization.  \\ \indent
For the real polarization $\Lambda_S$ generated by the 
solution $S$ of the Hamilton-Jacobi equation, the polarized 
sections take the form of the quantum wave functions  in the 
WKB approximation.  However, similar functions appear in the 
structure of the exact solution ${\sf f}_0$ for  the classical 
one-particle Liouville equation,  which becomes the Wigner 
transform ${\sf f}_\psi$ of $\psi$  if the configuration 
space is discretized.  Although ${\sf f}_\psi$ is not positive 
definite, it is integrable over $M$, and the phase-space 
overlap between two such functions  ${\sf f}_{\psi_1}, 
{\sf f}_{\psi_2}$ is proportional to $\vert \langle \psi_1 
\vert \psi_2 \rangle \vert^2$.
\\ \indent
 By discretization, ${\sf f}$  aquires a nonlocal character, as
 ${\sf f}_\psi ({\bf q},{\bf p})$ depends not only on $\vert 
\psi ({\bf q}) \vert^2$, but also on  $\bar{ \phi}_{\bf p} 
({\bf q}_-)\phi_{\bf p} ({\bf q}_+)$ with $\phi_{\bf p} 
({\bf q}) = \exp(- {\ri} {\bf q} \cdot {\bf  p} / \hbar ) 
\psi({\bf q})$  and ${\bf q}_\pm = {\bf q} \pm \hbar {\bf k} 
/2$  covering a whole domain of "quantum coherence", containing
 ${\bf q}$. The relationship between a possible lattice 
structure of the phase-space and the statistical interpretation
 of ${\sf f}_\psi$  remains  a  subject  worth of further 
consideration.  
\section{Appendix}
{\bf Definition 1.} $M$ is a complex manifold if it possesses 
an atlas $\{ (U_i,\varphi_i), i \in I \}$ where $U$ are open 
sets covering $M$, $\varphi_i: U_i \mapsto {\cal O}_i \subset 
{\mathbb C}^n$ is a diffeomorphism, and the transition 
functions $ c_{ij} = \varphi_j \circ \varphi _i^{-1} $ are 
holomorphic. If $p \in U_i \cap U_j$ then $ T_p  \varphi_i  
: T_pM \mapsto {\mathbb C}^n$, $ T_p  \varphi_j : T_pM \mapsto 
{\mathbb C}^n $, and $ T_p \varphi_i \circ (T_p \varphi_j)^{-1} 
\in {\rm GL}(n,{\mathbb C})$. \\
{\bf Definition 2.} Let $(M, \omega)$ be a complex symplectic 
manifold. Then $M$ is called a K\"ahler manifold if for any 
$ p \in M$ the complex structure $J_p \in {\rm Sp} (T_p M)$ 
and $\omega_p$ define a K\"ahler structure on $T_pM$,  
$ \omega_p (J_p x, J_p y) = \omega_p(x,y)$. $M$ is a positive 
K\"ahler manifold if $(x,y)_p \equiv \omega_p(x,J_p y)$ is 
positive definite.

\end{document}